\documentclass[structabstract]{aa}
\usepackage{txfonts}
\usepackage{graphicx}

\begin{document}

\title{The structure of thin accretion discs around magnetised stars}

     \author{Solomon Belay Tessema\inst{1,2}
          \and Ulf Torkelsson\inst{2}}
\institute{Department of Physics, Addis Ababa University,
     P.O.Box 1176, Addis Ababa, Ethiopia\\
     \email{newtonsolbel@yahoo.com}
     \and Department of Physics,  University of Gothenburg, SE 412 96 Gothenburg, Sweden\\
     \email{torkel@physics.gu.se}}
  \date{Received  /  Accepted  }
\abstract{}{We determine the steady-state of an axisymmetric thin accretion
disc with an internal dynamo around a magnetised star.}
{Starting from the vertically integrated equations of magnetohydrodynamics we
derive a single ordinary differential equation for a thin accretion disc around
a massive magnetic dipole and integrate this equation numerically from the outside inwards.}
{Our numerical solution shows that the torque between the star and the accretion
disc is dominated by the contribution from the dynamo in the disc.
The location of the inner edge of the accretion disc varies between
$R_{\rm A}$ and  $10R_{\rm A}$ depending mainly on the strength and direction of the magnetic field
generated by the dynamo in the disc}{}
\keywords{Accretion, accretion discs - Magnetohydrodynamics (MHD) -
Magnetic fields - Stars: neutron - X-rays: stars - Stars: pre-main sequence}

\maketitle

\section{Introduction}

In this paper we present a new solution for an accretion disc around
a magnetic star.  This star could be a neutron star, a white dwarf, or a
T Tauri-star, but we assume that it is a neutron star since it is easy
to measure the torque between the neutron star and the accretion disc by
timing the X-ray pulses from the neutron star.  The new feature of our
solution is to include the effect of an internal dynamo in the
accretion disc.  By doing this we hope to be able to explain the
torque reversals that have been observed in some X-ray pulsars.

\cite{shakura} formulated the standard model of a geometrically thin,
optically thick accretion disc.
They were able to obtain an
analytical solution of the height-integrated hydrodynamic equations,
after having introduced the
$\alpha$-prescription for the turbulent stress,
which transports the angular momentum outwards through the disc;
however, they did not explain why the disc is turbulent in the first place,
since a disc in Keplerian rotation is stable according to Rayleigh's criterion.
\cite{balbus1} instead showed that it is unstable if there is a
weak magnetic field in the disc.
Subsequent numerical simulations (e.g. \cite{hawley};  \cite{balbus})
confirmed that this
instability generates turbulence and that the resulting
turbulent stresses transport
angular momentum outwards.

The interaction between a magnetised  star and a surrounding accretion disc  is
one of the most poorly understood aspects of accretion. At the same time, it is
central for our understanding of the spin evolution of objects as diverse as
T Tauri stars and  X-ray pulsars.  The magnetic field of the star  penetrates
the surrounding accretion disc and couples the two. According to the
Ghosh \& Lamb (\cite{ghosh}) model, the part of the accretion disc that is located
inside the corotation radius provides a spin up torque on the star, since it
is rotating faster than the star, while the more slowly rotating outer part of
the accretion disc brakes the star. The net torque is determined by the
location of the inner edge of the disc, which moves inwards as the accretion
rate increases, thereby increasing the spin up-torque on the star.

\cite{campbell3} proposes  physical descriptions of the magnetic diffusivity
in terms of turbulence or buoyancy, and  \cite{campbell2}
solve the resulting equations numerically. For both forms of diffusivity, the
magnetic coupling between the disc and the star leads to an enhanced
dissipation in the inner part of the accretion disc compared to the standard
\cite{shakura} model. This raises the
temperature such that electron scattering dominates Kramer's opacity at
larger radii than is otherwise the case, thus increasing the fraction of the disc that is subject to the
(\cite{lightman}) instability.
\cite{brandenburg} considered a form of magnetic diffusivity that
allows further analytical progress to be made, but the qualitative results 
remain the same.

All the  models predict a positive correlation between the accretion rate and
the torque on the neutron star and even predict a negative torque on the
neutron star at very low accretion rates. Timing of X-ray pulsars during outbursts of Be/X-ray
transients have provided at least some qualitative support for such a
correlation (e.g. \cite{parmar}).
The BATSE instrument on the Compton Gamma
Ray Observatory made it possible to extend this database significantly
(\cite{bildsten}).  In particular there are a few X-ray pulsars with
permanent discs that are oscillating between phases of constant spin-up and
constant spin-down  without a significant difference in the
X-ray luminosity between these states, which appears to contradict the standard model for a disc-accreting X-ray pulsar.

\cite{nelson} propose that these torque-reversals can be the result of
transitions between co-rotating and counter-rotating accretion discs, though
several other models have also been proposed. \cite{torkelsson} argue that the
torque between an accreting star and its disc can be enhanced by the presence
of a magnetic field generated by the turbulence in the accretion disc. The
torque reversals are then the result of a reversal of the magnetic field
generated by this dynamo. However, he did not construct a self-consistent model
of the accretion disc. The aim of this paper is to construct such a model of an
accretion disc with an internal dynamo around a magnetic star. We work in
the spirit of
\cite{shakura} and assume that the disc is geometrically thin. In
Sect. 2 we start from the equations of magnetohydrodynamics
(MHD) and derive a single ordinary differential equation for the radial
structure of the accretion disc. We then present numerical solutions of this
equation in Sect. 3 and discuss the properties of these solutions in Sect.
4. Finally we summarise our conclusions in Sect. 5.

\section{Mathematical formulation}

We study a steady, thin axisymmetric Keplerian disc around a star
with a magnetic dipole field.
The basic equations describing the structure of the thin accretion disc can be
derived from the equations of magnetohydrodynamics.

\subsection{Conservation of mass}

In steady state the continuity equation takes the form
\begin{equation}\label{1}
    \mathbf{\nabla\cdot}\left(\rho \mathbf{v}\right) = 0
\end{equation}
where $\rho$ is the density and ${\bf v} = (v_{R},v_{\phi},v_{z})$ is the fluid velocity with radial, azimuthal, and vertical components , respectively.
For a thin axisymmetric disc and after neglecting a vertical outflow from the 
disc, we get
\begin{equation}\label{2}
    \frac{1}{R}\frac{\partial}{\partial R}\left(R\Sigma v_{R}\right)  = 0,
\end{equation}
where $\Sigma$ is the surface density
\begin{equation}\label{3}
    \Sigma = \int_{- H}^{H}\rho dz \simeq 2\rho H,
\end{equation}
and H is the halfthickness of the disc.
For a steady disc  the integral of Eq.(\ref{2}) gives
\begin{equation}\label{4}
    \dot{M} = -2\pi R\int_{-H}^{H}\rho v_{R}dz = -2\pi R\Sigma v_{R} = \mbox{constant},
\end{equation}
which is the accretion rate.

\subsection{Conservation of momentum}

If assuming a steady state  the Navier-Stoke's equation  can be written as
\begin{eqnarray}\label{5}
               \rho\left(\mathbf{v\cdot\nabla}\right)\mathbf{v}
               = -\mathbf{\nabla}P
               + \rho\mathbf{\nabla}\Phi +  \mathbf{J\times B}  \nonumber\\
               +   \mathbf{\nabla}\cdot\left(\rho\nu\left(\mathbf{\nabla v} -
               \frac{2}{3}\left(\mathbf{\nabla\cdot v}\right)\right)\right),
\end{eqnarray}
where  $P$ is  pressure, $\nu$   kinematic viscosity, $\Phi$  the gravitational potential
$\mathbf{J}= \frac{1}{\mu_{0}}\left(\mathbf{\nabla\times B}\right) = (J_R, J_\phi, J_z)$  the current
density, and $\mathbf{B} = (B_R, B_\phi, B_z)$  the  magnetic field. The 
viscosity is in general low, and we only  retain it where it plays a 
crucial role.

The radial component of Navier- Stoke's equation is
\begin{eqnarray}\label{6}
    \rho\left[ v_R\frac{\partial v_R}{\partial R} -
    \frac{v^{2}_\phi}{R}\right] =  \frac{B_\phi}{\mu_0}\left(\frac{\partial
B_R}{\partial z}-\frac{\partial B_z}{\partial R}\right)- \frac{B_z}{\mu_0}
\left(\frac{1}{R}\frac{\partial}{\partial R}(RB_\phi)\right)\nonumber\\
- \frac{\partial P}{\partial R} - \frac{\rho GM R}{(R^{2} + z^{2})^{3/2}}.
\end{eqnarray}
For a thin accretion disc $v_\phi\gg c_{\mathrm{s}}$ as shown below and
the dominant terms of the equation give us
\begin{equation}\label{7}
    v_\phi^{2} - \frac{GM}{R} = 0,
\end{equation}
which shows that the disc rotates in a Keplerian fashion.

In similar manner, the vertical component of the momentum equation for a
steady flow  is
\begin{eqnarray}\label{8}
    \rho \left[v_R \frac{\partial v_z}{\partial R} + v_z\frac{\partial v_z}{\partial
    z}\right] =  - \frac{B_\phi}{\mu_0}\frac{\partial B_\phi}{\partial z} -
    \frac{B_R}{\mu_0}\frac{\partial B_R}{\partial z} + \frac{B_R}{\mu_0}
    \frac{\partial B_z}{\partial R}\nonumber\\
    -  \frac{\partial P}{\partial z} -  \frac{\rho G M}{R^{2}}.
\end{eqnarray}
Neglecting  vertical outflows and assuming the magnetic field to be weak the 
equation reduces to the equation of hydrostatic equilibrium
\begin{equation}\label{9}
   \frac{1}{\rho} \frac{\partial P}{\partial z}= -\frac{GM}{R^{2}}\frac{z}{R}.
\end{equation}
Using $H$ as the halfthickness of the disc, the pressure at the midplane
of the disc is
\begin{equation}\label{10}
        P = \frac{1}{2}H\Sigma\frac{GM}{R^{3}},
\end{equation}
but the hydrostatic equilibrium can also be expressed as
\begin{equation}\label{11}
        \frac{H}{R} = \frac{c_{\mathrm s}}{v_{\mathrm {kepl}}},
\end{equation}
which shows that the Keplerian velocity
is  highly supersonic in a thin accretion disc, as assumed above.

The azimuthal component of Navier-Stoke's equation
reduces to
\begin{eqnarray}\label{12}
    \rho\left(\frac{v_{R}}{R}\frac{\partial}{\partial
    R}\left(Rv_\phi\right)\right) = \frac{B_R}{\mu_0}\frac{1}{R}
    \frac{\partial}{\partial R}(RB_\phi) + \frac{B_z}{\mu_0}
    \frac{\partial B_\phi}{\partial z}    \nonumber\\+
    \frac{1}{R^{2}}\frac{\partial}{\partial R}
    \left[R^{3}\rho\nu\frac{\partial}{\partial R}
    \left(\frac{v_\phi}{R}\right)\right].
\end{eqnarray}
We  neglect \,$\frac{B_R}{R}\frac{\partial}{\partial R}(R B_\phi)$   because
the the radial length scale is much longer than the vertical length scale
in a thin accretion disc. 
Integrating Eq.  (\ref{12}) vertically across the disc and multiplying both sides by $R$, we get
\begin{equation}\label{14}
    \Sigma \left(v_R\frac{\partial l}{\partial R}\right) =
     \left[\frac{B_z B_\phi}{\mu_0}\right]_{-H}^{H}R  + \frac{1}{R}
     \frac{\partial}{\partial R}\left[R^{3}\nu\Sigma\frac{\partial}{\partial R}
    \left(\frac{l}{R^{2}}\right)\right],
\end{equation}
where  the specific angular momentum $l = Rv_\phi \propto R^{1/2}.$
The magnetic term describes the exchange of angular momentum between the disc
and the star via the magnetosphere.
This term vanishes if $B_\phi$ is an even function of $z$, but the shear 
between the disc and the stellar magnetosphere generates an odd $B_\phi$ whose 
value in the upper half of the disc is
\begin{equation}\label{15}
B_{\phi,{\rm shear}} = -\gamma B_z
\frac{\Omega_{\mathrm{k}}- \Omega_{\mathrm{s}}}{\Omega_{\mathrm{k}}},
\end{equation}
where $\Omega_{\rm k} = v_\phi/R$,  $\Omega_{\mathrm {s}}$ is the angular
velocity of the star, and $\gamma$ is a dimensionless parameter
of a few (Ghosh \& Lamb \cite{ghosh}).

In this paper we consider the effect of adding
a large-scale toroidal field
that is generated by an internal dynamo in the accretion disc. Such
a dynamo is a natural consequence of the magnetohydrodynamic turbulence in the
accretion disc (e.g. \cite{balbus}). To estimate the size of
$B_{\phi,{\rm dyn}}$, we assume for the moment that the viscous stress in the accretion disc is due to the internal magnetic stress
\begin{equation}\label{16}
f_{R\phi} = \frac{B_R B_{\phi,{\rm dyn}}}{\mu_0} = \alpha_{\mathrm{ss}}P,
\end{equation}
where we use the \cite{shakura} prescription for the viscosity in the last
equality.
Based on the results of numerical simulations of magnetohydrodynamic turbulence
in accretion discs (e.g. Brandenburg et al. 1995)  Torkelsson (1998) argues 
that
\begin{equation}\label{17}
\gamma_{\mathrm {dyn}} = \frac{B_\phi}{B_R} \sim \frac{B_\phi}{B_z}
\end{equation}
where $\gamma_{\rm dyn} \sim 10$.  However, this $B_\phi$ is the sum of the
large-scale field and a small-scale turbulent field, that is also contributing
to the stress $f_{R\phi}$ through its correlation with a turbulent $B_R$-field.
Since the large-scale field might be a small fraction of the total field
we multiply $B_\phi$ with a factor $\epsilon$ to get an estimate for
$B_{\phi,{\rm dyn}}:$
\begin{equation}\label{B-tor}
       B_{\phi,{\rm dyn}} = \epsilon
\left(\alpha_{\rm ss}\mu_{0}\gamma_{\rm dyn}
P\right)^{1/2},
\end{equation}
where
$-1 \leq \epsilon \leq1$,
and a negative value describes a magnetic field which is pointing in the negative
$\phi$-direction at the upper disc surface. 

We can now estimate the magnetic pressure in the accretion disc, to which
the toroidal magnetic field is the main contributor. According
to Eq. (\ref{B-tor}), the
pressure is approximately
\begin{equation}
  \frac{B_{\phi, {\rm dyn}}^2}{2\mu_0} = \frac{1}{2} \epsilon^2 \alpha_{\rm ss}
\gamma_{\rm dyn} P.
\end{equation}
Since we use $|\epsilon| \le 1$, $\alpha_{\rm ss} = 0.01$ and 
$\gamma_{\rm dyn} = 10$ in our models, 
we see that the magnetic pressure will not 
significantly affect the vertical structure of the accretion disc.

The vertical magnetic field can likewise be split up into two 
components; (i) the stellar dipolar magnetic field, whose value in the stellar
equatorial plane is
\begin{equation}
  B_{z, {\rm dipole}} = - \frac{\mu}{R^3},
\end{equation}
where $\mu$ is the magnetic dipole moment; and (ii) a dynamo component 
$B_{z, {\rm dyn}}$.
There is no obvious way to model $B_{z,{\rm dyn}}$ within our one-dimensional
model, but numerical simulations like those by Brandenburg et al. 
(\cite{brandenburg2}) suggest that $B_z$ and $B_R$ are comparable, so
we expect that $B_{z,{\rm dyn}} \sim B_{\phi,{\rm dyn}}/\gamma_{\rm dyn}$.
In fact, $B_{z,{\rm dyn}}$ will modify the structure of the poloidal
magnetic field, but this can happen even if there is no internal dynamo in the 
disc because of the currents that are induced in the disc (e.g. Bardou \& 
Heyvaerts \cite{bardou}).

We can now expand the product $B_z B_\phi$ as
\begin{eqnarray}
  B_z B_\phi =\left(B_{z, {\rm dipole}} + B_{z, {\rm dyn}}\right) 
\left(B_{\phi, {\rm shear}} + B_{\phi, {\rm dyn}}\right) =
\nonumber \\
B_{z, {\rm dipole}} B_{\phi, {\rm shear}} + 
B_{z, {\rm dipole}} B_{\phi, {\rm dyn}} +
\nonumber \\
B_{z, {\rm dyn}} B_{\phi, {\rm shear}} +
B_{z, {\rm dyn}} B_{\phi, {\rm dyn}}.
\end{eqnarray}
As we see below the term $B_{z, {\rm dipole}} B_{\phi, {\rm dyn}}$ is
significantly greater than $B_{z, {\rm dipole}} B_{\phi, {\rm shear}}$ almost
everywhere in the disc.  The term 
$B_{z, {\rm dyn}} B_{\phi, {\rm shear}}$ is smaller in size than 
$B_{z, {\rm dipole}} B_{\phi, {\rm dyn}}$ by a factor 
$\gamma/\gamma_{\rm dyn}$, and since
it is difficult to model it, we ignore it.
Finally the term $B_{z, {\rm dyn}} B_{\phi, {\rm dyn}}$ does not contribute
directly to the exchange of angular momentum between the accretion disc and the
accretor, though it does affect it indirectly 
by contributing to the radial transport of angular momentum 
through the disc.
Although this term can be important, we have decided to ignore it since there
is no obvious way to model it in our one-dimensional approach.  One should
notice here that this does not change the main qualitative conclusion of
this paper that an internal dynamo in the accretion disc makes a significant
contribution to the exchange of angular momentum between the disc and the 
star; on the contrary, we would see a stronger effect if we were to keep the 
$B_{z, {\rm dyn}} B_{\phi, {\rm dyn}}$-term.
We now expand the $B_z B_\phi/\mu_0$ term
in Eq. (\ref{14})
\begin{eqnarray}\label{19}
    \Sigma \left(v_R\frac{\mbox{d} l}{\mbox{d} R}\right) =
     2\frac{B_z}{\mu_0} \epsilon
    \left(\alpha_{\mathrm{ss}}\mu_0\gamma_{\mathrm{dyn}}P\right)^{1/2}R  -
    2\gamma\frac{B_z^2}{\mu_0}\frac{\Omega_{\mathrm{k}}- 
    \Omega_{\mathrm{s}}}{\Omega_{\mathrm{k}}}R\nonumber\\
      + \frac{1}{R}\frac{\mbox{d}}{\mbox{d} R}\left[R^{3}\nu\Sigma
    \frac{\mbox{d}}{\mbox{d} R}
    \left(\frac{l}{R^{2}}\right)\right].
\end{eqnarray}

\subsection{Conservation of energy}

For a slow inflow of matter through an optically thick disc,
the local viscous dissipation  $\mathbf{v\cdot f_\nu}$ is balanced by the
radiative losses  $\mathbf{\nabla\cdot F_{\mathrm{rad}}}.$
This gives us
\begin{equation}\label{20}
    \frac{9}{8}\nu\Sigma\frac{GM}{R^{3}} = \frac{4\sigma T^{4}_{\mathrm{c}}}{3\tau}
\end{equation}
where  $T_c$ is the temperature at the midplane of the disc, and
$\sigma$ the Stefan-Boltzmann constant. The optical depth of the disc is given by
\begin{equation}\label{21}
    \tau = \rho H\kappa_{\mathrm{R}} = \frac{1}{2}\Sigma\kappa_{\mathrm{R}},
\end{equation}
and we assume that the opacity is given by  Kramer's law
\begin{equation}\label{22}
\kappa_{\mathrm{R}} = \kappa_0 \rho T_{\mathrm{c}}^{-7/2}
\mathrm{m^{2}\, kg^{-1}},
\end{equation}
where $\kappa_0 = 5\times 10^{20}$\,m$^5$ $\rm kg^{-2}$\,K$^{-7/2}$.

Equation (\ref{20}) should also contain a term describing the
magnetic dissipation
\begin{equation}
  2H \frac{J^2}{\sigma} = \frac{2B_\phi^2}{\sigma \mu_0^2 H} = 
\frac{2\epsilon^2 \alpha_{\rm ss} \gamma_{\rm dyn} \eta P}{H},
\end{equation}
where $\eta = 1/\sigma \mu_0$ is the magnetic
diffusivity, and we have used Eq. (\ref{B-tor}) in the last equality.  
This should be compared to
\begin{equation}
  \frac{9}{8} \nu \Sigma \frac{GM}{R^3} = \frac{9}{4} \frac{\nu P}{H},
\end{equation}
so we see that the magnetic dissipation is negligible as long as $\eta \approx
\nu$.

\subsection{Structure equations}

We reduce these equations to a single ordinary differential equation for the 
radial structure of the accretion disc. First we assume the equation of state 
of an ideal gas,
\begin{equation}\label{23}
   P = \frac{\rho k_{\mathrm{B}}T_{\mathrm{c}}}{m_{\mathrm{p}}\bar{\mu}},
\end{equation}
where  $k_{\mathrm{B}}$ is the
Boltzmann constant,  $\bar{\mu}$  the mean molecular weight, and  
$m_{\mathrm{p}}$ the mass of a proton,
but the pressure can also be expressed using the equation of hydrostatic 
equilibrium
\begin{equation}\label{24}
\frac{1}{2}\frac{\Sigma G M H}{R^{3}} = \frac{\rho k_{\mathrm{B}}T_{\mathrm{c}}}{\bar{\mu}m_{\mathrm{p}}},
\end{equation}
which gives us a relation between $H$ and $T_{\rm c}$
\begin{equation}\label{25}
  H =  \left( \frac{ k_{B}}{m_{\mathrm{p}}\bar{\mu}GM}\right)^{1/2}
T_{\mathrm{c}}^{1/2}R^{3/2}.
\end{equation}
The viscous stress tensor
gives us the equation
\begin{equation}\label{26}
     f_{r\phi}= \frac{3}{4}\Sigma\nu \left(\frac{GM}{R^{3}}\right)^{1/2}H^{-1}
=  \alpha_{\mathrm{ss}}P(r),
\end{equation}
which we solve for the density of the gas
\begin{equation}\label{27}
   \rho = \frac{3}{4}\alpha_{\mathrm{ss}}^{-1}
\left( \frac{m_{\mathrm{p}}\bar{\mu}}{k_{\mathrm{B}}}\right)^{3/2}
\frac{\nu\Sigma}{T_{c}^{3/2}R^{3}}GM.
\end{equation}
The optical depth of the disc is
\begin{equation}\label{28}
    \tau = \kappa_{\mathrm{0}}\frac{9}{16}\alpha_{\mathrm{ss}}^{-2}
    \left(\frac{m_{\mathrm{p}}\bar{\mu}}{k_{\mathrm{B}}}\right)^{5/2}
    (GM)^{3/2}(\nu\Sigma)^{2}T_{c}^{-6}R^{-9/2}.
\end{equation}

Using Eqs. (\ref{20}) and (\ref{28})  we get
\begin{equation}\label{29}
    T_{\mathrm{c}}= C\bar{\mu}^{1/4}\alpha_{\mathrm{ss}}^{-1/5}M^{1/4}
y^{3/10}R^{-3/4},
\end{equation}
where $y = \nu\Sigma$ and
\begin{equation}\label{30}
C=\left(\frac{243\kappa_{\mathrm{0}}}{512\sigma} \right)^{1/10}
\left(\frac{Gm_{p}}{k_{B}}\right)^{1/4}.
\end{equation}
The pressure is then given by
\begin{equation}\label{31}
    P = C_{1}\bar{\mu}^{3/8}\alpha_{ss}^{-9/10}M^{7/8}y^{17/20}R^{-21/8},
\end{equation}
where
\begin{equation}\label{32}
C_{1}=\frac{3}{4} G^{1/2}
\left(\frac{243\kappa_{\mathrm{0}}}{512\sigma} \right)^{-1/20}
\left(\frac{Gm_{p}}{k_{B}}\right)^{3/8}.
\end{equation}
The $\phi$-component of the magnetic field generated by the internal dynamo
can be expressed using Eqs. (\ref{B-tor}) and (\ref{31}) as
\begin{equation}\label{33}
    B_{\phi,{\rm dyn}} = C_{2}\epsilon
\gamma_{\mathrm{dyn}}^{1/2}\alpha_{\mathrm{ss}}^{1/20} \bar{\mu}^{3/16}
M^{7/16}y^{17/40}R^{-21/16},
\end{equation}
where
\begin{equation}\label{34}
C_{2} = \left(\mu_0 C_1\right)^{1/2}.
\end{equation}
The magnetic field due to the shear can be written as:
\begin{equation}\label{35}
    B_{\phi,{\rm shear}} = \frac{\mu\gamma}{R^{3}}
    \left[1- \left(\frac{R}{R_{\mathrm{c}}}\right)^{3/2}\right],
\end{equation}
where
\begin{equation}\label{36}
   R_{c}= \left(\frac{GMP_{\rm spin}^{2}}{4\pi^{2}}\right)^{1/3}\simeq 1.5\times 10^{6}
P_{\rm spin}^{2/3}M_{1}^{1/3}\mbox{m}
\end{equation}
is the corotation radius, at which the Keplerian angular velocity is the same
as the stellar angular velocity.
Here $P_{\rm spin} = 2\pi/\Omega_{\mathrm{s}}$ is the spin period of the star,
and $M_1 = M/M_\odot$.
Equations (\ref{33}) and (\ref{35}) show that the magnetic field generated by the internal dynamo
varies more slowly with  radius $\sim R^{\mathrm{-1.3}} $ than the magnetic
field due to shear, $\sim$ $R^{\mathrm{-3}}$. Thus the
dynamo component dominates at large radii.

Equation (\ref{19}) gives us an ordinary differential equation for $y$
\begin{eqnarray}\label{37}
     y' = \frac{\dot{M}}{6\pi R} - \frac{y}{2 R} - C_{3} y^{17/40}R^{-45/16}
     \nonumber \\
    - C_{4}R^{-9/2}\left[1 - \left(\frac{R}{R_{c}}\right)^{3/2}\right],
\end{eqnarray}
where
\begin{equation}\label{38}
   C_{3} = \frac{4}{3\mu_{0}\sqrt{G}} C_{2}\epsilon
\gamma_{\mathrm{dyn}}^{1/2}\alpha_{\mathrm{ss}}^{1/20}M^{-1/16} \bar{\mu}^{3/16}
\mu
\end{equation}
and
\begin{equation}\label{39}
C_{4} = \frac{4\gamma}{3\mu_0\sqrt{G}}M^{-1/2} \mu^2.
\end{equation}
The solution of Eq. (\ref{37}) approaches the Shakura-Sunyaev solution at large
radii, thus giving us the boundary condition
$y  \longrightarrow \frac{\dot{M}}{3\pi}$  as  $R\longrightarrow\infty.$

We introduce the dimensionless variable $\Lambda$  through
\begin{equation}\label{40}
    y = \Lambda\dot{M}
\end{equation}
and a dimensionless radial coordinate through
\begin{equation}\label{41}
    R = rR_\mathrm{{A}},
\end{equation}
where $ R_\mathrm{{A}}$ is the Alfv\'en radius,
\begin{equation}\label{42}
    R_{A} =\left(\frac{2\pi^{2}\mu^{4}}{GM\dot{M}^{2}\mu_{0}^{2}}\right)^{1/7} =  5.1\times 10^{6} {\dot{M}_{13}}^{-2/7}M_{1}^{-1/7}\mu_{20}^{4/7}\mbox{m},
\end{equation}
which is given by putting the magnetic pressure equal to the ram pressure of the accreting fluid
(e.g. \cite{frank}).
Here $\dot{M}_{\mathrm{13}}$ represents the mass transfer rate in units of
$10^{13}$\,kg\,s$^{-1}$, and $\mu_{\mathrm{20}}$ is the stellar magnetic
dipole moment
in units of $10^{20}$\,T\,m$^{3}$. The boundary condition is then
$\Lambda\longrightarrow \frac{1}{3\pi}$ as $r\longrightarrow\infty$, and Eq.
(\ref{37}) can be written as
\begin{eqnarray}\label{43}
    \Lambda' = -\frac{\Lambda}{2 r}  +  \frac{1}{6\pi r} -
C_{5}\Lambda^{17/40}r^{-45/16} - C_6 r^{-9/2}(1 - \omega_{s}r^{3/2}),
\end{eqnarray}
where
\begin{equation}\label{44}
  C_{5} = 4.9\epsilon\alpha_{\mathrm{ss}}^{1/20}\gamma_{\mathrm{dyn}}^{1/2}
\bar{\mu}^{3/16}M_{\mathrm{1}}^{11/56}\dot{M}_{13}^{-2/35}\mu_{20}^{-1/28},
\end{equation}
\begin{displaymath}
C_{6} = 0.3 \gamma,
\end{displaymath}
and
\begin{equation}\label{45}
   \omega_{\rm s} = \left(\frac{R_{\mathrm{A}}}{R_{\mathrm{c}}} \right)^{3/2} =
6.3\,\dot{M}_{\mathrm{13}}^{-3/7}M_{\mathrm{1}}^{-5/7}\mu_{\mathrm{20}}^{6/7}P^{\mathrm{-1}}
\end{equation}
is the fastness parameter.

\begin{table}
\caption{Spin parameters of the models}
\label{table:1}
\begin{tabular}{lll}
\hline
   $P$\,(s) & $R_{\rm c}$\,(m) & $\omega_{\rm s}$ \\
\hline
   7 & $6.1\times 10^{6}$ & 0.71 \\
   18.7 & $1.2\times 10^{7}$ & 0.26 \\
   100 &  $3.6\times 10^{7}$ & 0.05 \\
\hline
\end{tabular}
\end{table}

\section{Numerical solution}

\subsection{Global solutions}

We integrate Eq. (\ref{43}) inwards from a large radius, usually $100R_{\mathrm{A}}$,
to which we impose the boundary condition that $\Lambda =  \frac{1}{3\pi}$,
which is similar to the approach by \cite{brandenburg},  but most studies
of accretion discs have rather applied a boundary condition
at the inner edge of the disc, and then integrated the equations outwards.

There are two possible boundary conditions that can be applied at the inner
edge of the accretion disc, either that $\Lambda = 0$, which corresponds
to that $\rho$ and $T \to 0$ at the inner edge (case D), or that
\begin{equation}
  \frac{\mbox{d}}{\mbox{d}r} \left(r^3\Lambda \frac{\mbox{d}\Omega}{\mbox{d}r}
\right) = 0,
\end{equation}
which means that the viscosity does not at all contribute to driving the
accretion at this radius.
Case D is the boundary condition that has been
most widely used and that was adopted for instance by \cite{shakura}.  In this
case the density drops to zero because the inflow velocity becomes infinite,
which is of course not realistic, and a more accurate treatment shows that
the inflow becomes transonic close to this position \cite{paczynski}.
In case V the inflow at the inner edge of the accretion disc is driven completely
by the transfer of excess angular momentum from the accreting matter to the stellar
magnetic field (e.g. \cite{wang95}).

For our fiducial model, we take a neutron star of $M= 1.4M_{\mathrm{\odot}}$
and a magnetic moment of $10^{20}$\,T\,m$^3$, which is accreting at
$10^{\mathrm{13}}\,\mathrm{kg\,s^{-1}}$.
The dimensionless parameters $\gamma$ and $\gamma_{\rm dyn}$ are, respectively,
1 and 10 in
our fiducial model, while $\alpha_{\rm ss} = 0.01$.  The exact values of
$\alpha_{\rm ss}$ and $\gamma_{\rm dyn}$ are  unimportant, since we
vary the parameter $\epsilon$ below, but $\gamma$ influences the solutions
in its own way as is shown at the end of this section.
We consider three different spin
periods with corresponding
corotation radii and fastness parameters (see Table.1). The system
goes into the propeller regime for $\omega_s\geq 1$ (e.g. \cite{illarionov},
Ghosh \& Lamb 1979).

\begin{figure}[tbhp]
\centerline{\includegraphics[width = 8.5cm]{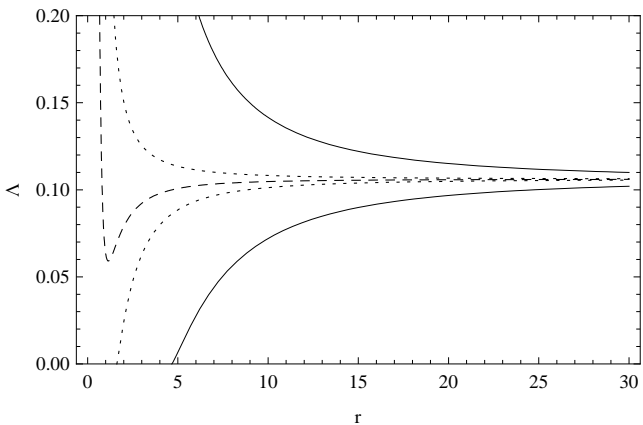}}
\caption{ $\Lambda(r)$ for our fiducial neutron star with a spin period of 7\,s
and $\epsilon= 1, 0.1, 0, -0.1, -1$ from the top to the bottom.
}
\label{Lambda1}
\end{figure}

\begin{figure}[tbhp]
\includegraphics[width = 8.5cm]{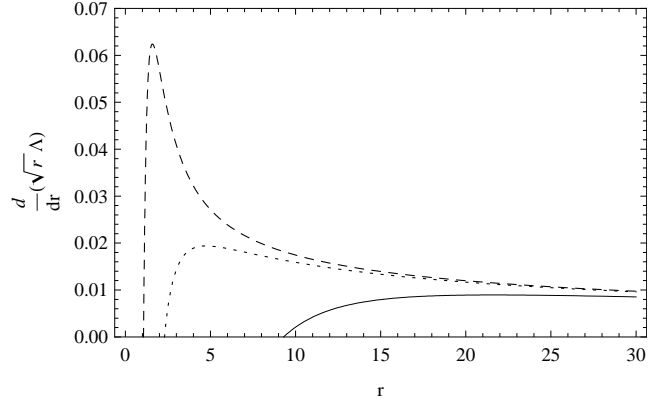}
\caption{$\frac{d}{dr}\left(\sqrt{r}\Lambda\right)$, as a function of $r$ for
the fiducial neutron star with a spin peroid of 7\,s
and with $\epsilon = 1,\, 0.1$, and $0$ from the bottom to the top.}
\label{innerradius}
\end{figure}

Firstly, for a spin period of 7\,s and $\epsilon = 1, \, 0.1, \,0, \, - 0.1,
\,$ and $ - 1$,
we get the solutions shown from the top to the bottom of Fig. \ref{Lambda1}.
The $\epsilon = -1$ and $-0.1$ solutions have case D inner boundaries at
4.7 and 1.6 $R_{\rm A}$, respectively, while the other three
solutions have case V inner boundaries (Fig. \ref{innerradius}).
The inner boundary is close to $R_{\rm A}$ if
$\epsilon = 0,$  which corresponds to the absence of an internal disc dynamo, but moves outwards
as $\left|\epsilon\right|$\, increases.
In case V, the solution continues inside
the inner edge of the accretion disc, but the viscosity counteracts the accretion,
which is instead driven by the magnetic stresses. This regime has been
described by \cite{campbell}, who discusses how the disc is disrupted in this
region, we assume that this region belongs to  a boundary layer
that we do not attempt to model in this paper.
On the other hand, all solutions approach the Shakura-Sunyaev solution at large radii, as
required by our boundary condition.

\begin{figure}[tbhp]
\centerline{\includegraphics[width = 8.5cm]{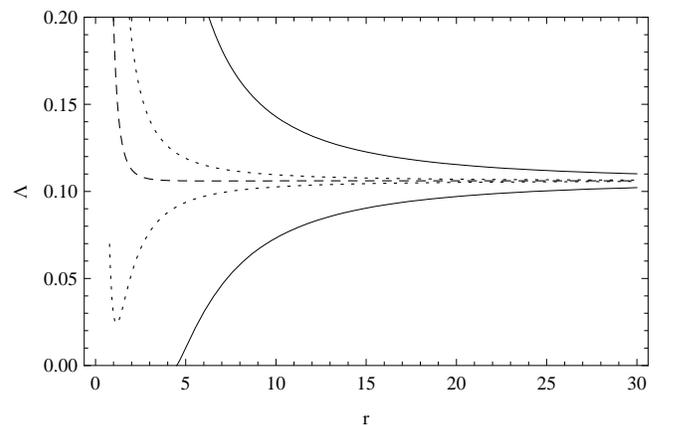}}
\caption{ $\Lambda(r)$ for our fiducial neutron star with a spin period of
100 s and  $\epsilon= 1,\,0.1,\,0,\,-0.1,\,-1$ from the top to the bottom.
}
\label{Lambda2}
\end{figure}

\begin{figure}[tbhp]
\centerline{\includegraphics[width = 8.5cm]{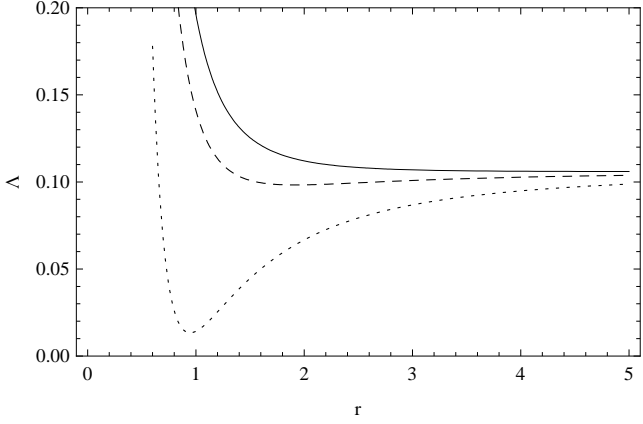}}
\caption{$\Lambda(r)$ for our fiducial neutron star with $\epsilon=$ 0 and
spin periods of 100, 18.7, and 7\,s from the top to the bottom.
}
\label{Lambda3}
\end{figure}

\begin{figure}[tbhp]
\centerline{\includegraphics[width = 8.5cm]{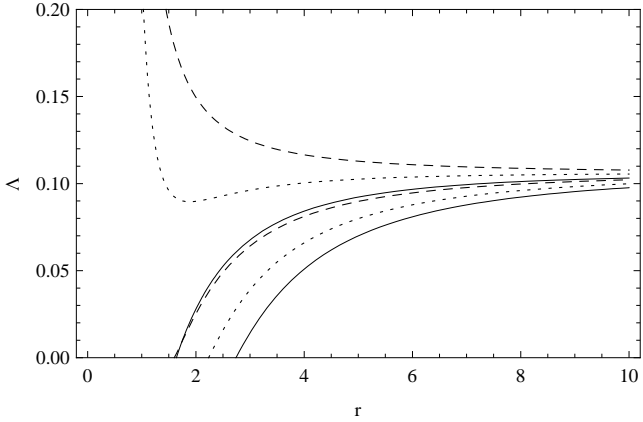}}
\caption{$\Lambda(r)$ for our fiducial neutron star with spin period of 7\,s and $\epsilon= 0.1$ and
$\gamma = 1,\,3,$ and 5 ( upper dashed, dotted, and solid lines, respectively), $\epsilon= -0.1$ and
$\gamma = 1,\,3$, and 5 ( lower dashed, dotted, and solid lines, respectively).}
\label{Lambdagamma1}
\end{figure}

By increasing the spin peroid to 100\,s, we see that the $\epsilon = -0.1$
solution also develops a case V inner boundary (Fig. \ref{Lambda2}),
but there are only small quantitative
changes for the $\epsilon = -1$ and -1 solutions. (The inner boundary
of the $\epsilon = -1$ solution is now located at $4.4 R_{\rm A}$.)
The dependence of the solution on the spin period can be better seen in
Fig. \ref{Lambda3}, where we vary the spin period, when $\epsilon$ is fixed
to 0.  Then $\Lambda$ has a local minimum  for $P_{\rm s} = 7$\,s,  but this minimum
weakens and disappears as the spin period is increased.  This is similar to
how $\Lambda$ depends on $\epsilon$.  For $\epsilon = -1,$   $\Lambda$ has an
unphysical negative local minima at a small radius, such that the physical
solution has a case D boundary at several Alfv\'en radii.  As $\epsilon$ is
increased, the local minimum grows and the solution develops a case V inner boundary when the
minimum becomes positive.  Increasing $\epsilon$ removes the
local minimum further and the solution becomes a strictly decreasing function of $r$,
which asymptotically approaches $1/3\pi$ as required by our boundary
condition.
Increasing $\gamma$ has a similar effect to decreasing $\epsilon$ (Fig \ref{Lambdagamma1}).

\subsection{The structure of an accretion disc}

The physical structure of the accretion disc can now be expressed
using $\Lambda$  and $r= R/R_{A}.$ We have that
\begin{equation}\label{47}
   \Sigma = 3.7\times 10^3\alpha_\mathrm{ss}^{-4/5}M_{\mathrm{1}}^{5/14}
\dot{M}_{13}^{32/35}\mu_{20}^{-3/7}\Lambda(r)^{7/10}r^{-3/4}\,\mathrm{kg\,m^{-2}}
\end{equation}

\begin{equation}\label{48}
  T_{\rm c}= 4.8\times 10^{5}\alpha_\mathrm{ss}^{-1/5} \bar{\mu}^{1/4}M_{1}^{5/14}
\dot{M}_{13}^{18/35}\mu_{20}^{-3/7}\Lambda(r)^{3/10}r^{-3/4}\,\mathrm{K}
  \end{equation}

\begin{equation}\label{49}
\frac{H}{R}= 0.012\alpha_\mathrm{ss}^{-1/10} \bar{\mu}^{-3/8}
M_{1}^{-11/28}\dot{M}_{13}^{4/35}\mu_{20}^{1/14}\Lambda(r)^{3/20}r^{1/8}
\end{equation}

\begin{eqnarray}\label{50}
   \rho = 3.0\times 10^{-2}\alpha_\mathrm{ss}^{-7/10} \bar{\mu}^{9/8}
M_{1}^{25/28}
\dot{M}_{13}^{38/35}\mu_{20}^{-15/14}\Lambda(r)^{11/20}r^{-15/8}\nonumber\\
\,\mathrm{kg\,m^{-3}}
\end{eqnarray}

\begin{equation}\label{51}
   \tau =  370\alpha_\mathrm{ss}^{-4/5}\bar{\mu}
\dot{M}_{13}^{1/5}\Lambda(r)^{1/5}
\end{equation}

\begin{equation}\label{52}
    \nu = 2.7\times 10^{9}\alpha_\mathrm{ss}^{4/5} \bar{\mu}^{-3/4}
M_{1}^{-5/14}
\dot{M}_{13}^{3/35}\mu_{20}^{3/7}\Lambda(r)^{3/10}r^{3/4}\,\mathrm{m^{2}\,s^{-1}}
\end{equation}

\begin{equation}\label{53}
    v_{R} = 84\alpha_\mathrm{ss}^{4/5} \bar{\mu}^{-3/4}M_{1}^{-3/14}
\dot{M}_{13}^{13/35}\mu_{20}^{-1/7}\Lambda(r)^{-7/10}r^{-1/4} \,\mathrm{m\,s^{-1}}
\end{equation}

\begin{equation}\label{54}
    B_\mathrm{\phi,{dyn}} = 12 \epsilon \gamma_{\rm dyn}^{1/2}
\alpha_\mathrm{ss}^{1/20} \bar{\mu}^{3/16}M_{1}^{5/8}
\dot{M}_{13}^{4/5}\mu_{20}^{-3/4}\Lambda(r)^{17/40}r^{-21/16} \,\mathrm{T}
\end{equation}

\begin{equation}\label{55}
    B_\mathrm{\phi,{shear}} = 0.75 \gamma M_{1}^{3/7}\dot{M}_{13}^{6/7}
\mu_{20}^{-5/7}r^{-3}\left(1-\omega_{s}r^{3/2}\right) \,\mathrm{T}.
\end{equation}

\begin{figure}[tbhp]
\centerline{\includegraphics[width = 8.5cm]{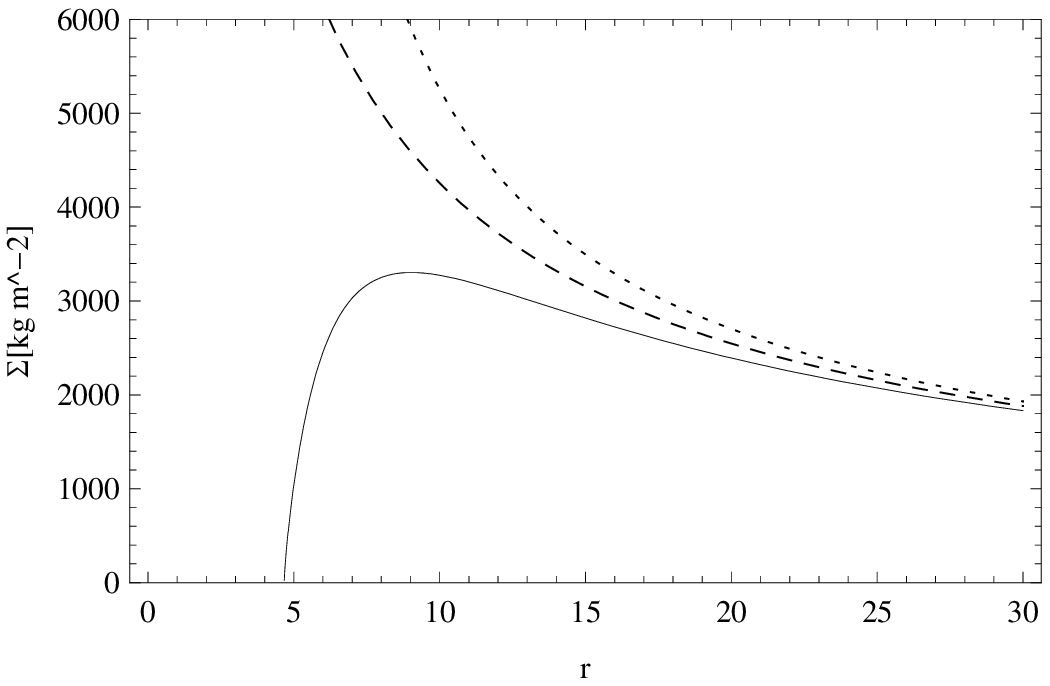}}
\caption{ $\Sigma$  as a function of $r$ for the fiducial neutron star with a spin period of 7\,s
and $\epsilon= -1$ (solid line),
$\epsilon= 0 $ (dashed line), and $\epsilon= 1$ (dotted line).}\label{Sigma1}
\end{figure}

\begin{figure}[tbhp]
\centerline{\includegraphics[width = 8.5cm]{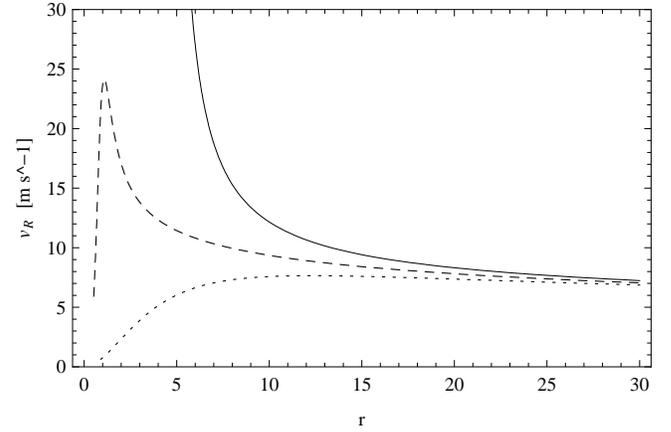}}
\caption{ $V_{R}$  as a function of $r$ for the fiducial neutron star with
a spin period of 7\,s
and $\epsilon= -1$ (solid line),
$\epsilon= 0 $ (dashed line), and $\epsilon= 1$ (dotted line).}\label{radialvelocity}
\end{figure}

\begin{figure}[tbhp]
\centerline{\includegraphics[width = 8.5cm]{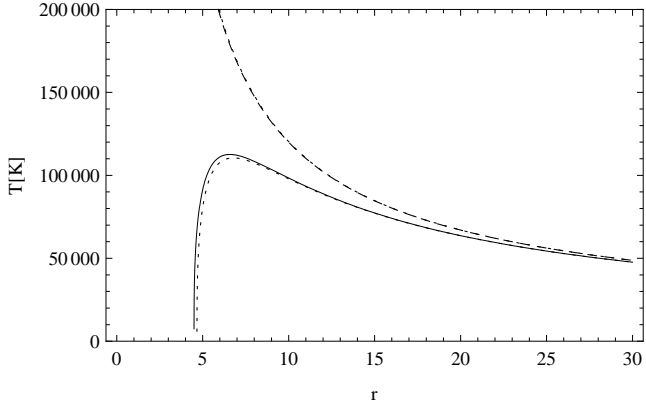}}
\caption{ $T_{\mathrm{c}}$ as a function of $r$.
The two lower curves show discs with $\epsilon = -1$ around neutron stars
with spin periods of 7\,s (dashed line) and 100\,s (solid line), respectively.
The two upper curves show discs with $\epsilon = 1$ around neutron stars with
spin periods of 7\,s (dashed line) and 100\,s (dotted line), respectively.}
\label{Temp1}
\end{figure}

In Fig. \ref{Sigma1} we show the surface density as a function of radius for
the fiducial neutron star with a spin peroid of 7\,s and
$\epsilon =  -1$, 0, and 1, respectively.  For $\epsilon = -1$ the
surface density
attains a local maximum, while the other models have surface densities that
are strictly decreasing functions of $r$.
Figure \ref{radialvelocity} shows the radial velocity as a function of $r$. 
Since $v_{R}\propto \Sigma^{-1}$, it becomes infinite as the inner edge of the 
disc for $\epsilon = -1$, but it stays finite for  $\epsilon = 0$ and 1 and 
eventually goes to 0 in the boundary layer.  Increasing the spin period of the 
neutron  star has a very marginal effect on the accretion disc, though 
$T_{\rm c}$, which is proportional to $\Sigma^2$,  increases somewhat 
(Fig.\ref{Temp1}).

\begin{figure}[tbhp]
\centerline{\includegraphics[width = 8.5cm]{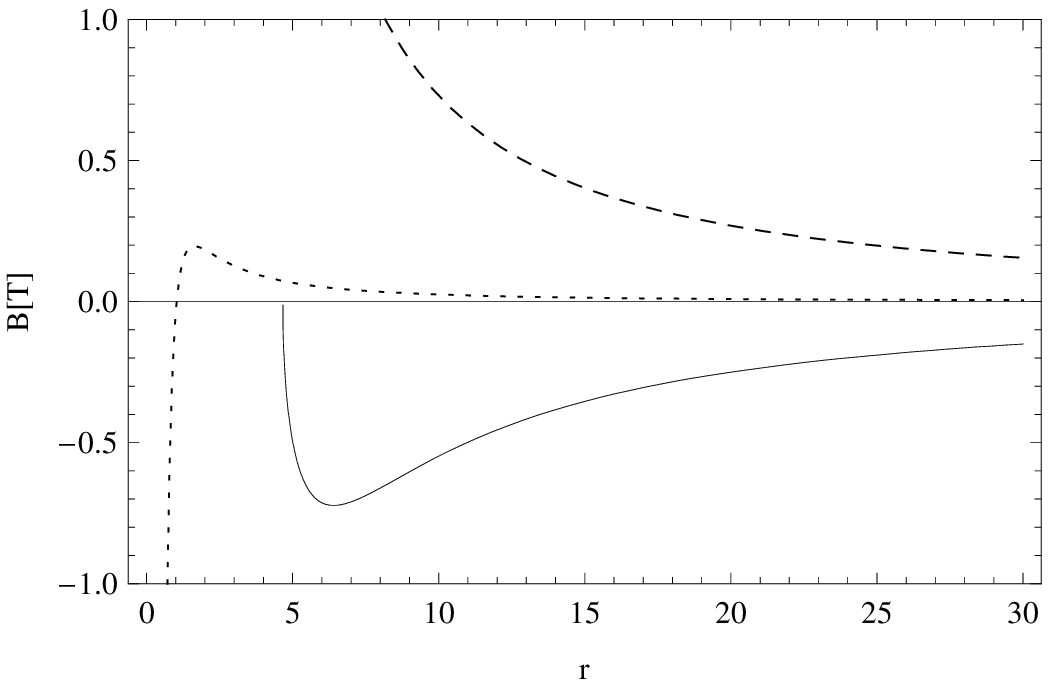}}
\caption{The toroidal magnetic field for the fiducial neutron star with a
spin period of 7\,s.  The solid and dashed lines show the field generated by
the dynamo for $\epsilon = -1$ and 1, respectively, while the dotted line shows
the magnetic field generated by the shear}
\label{toroidal1}
\end{figure}

\begin{figure}[tbhp]
\centerline{\includegraphics[width = 8.5cm]{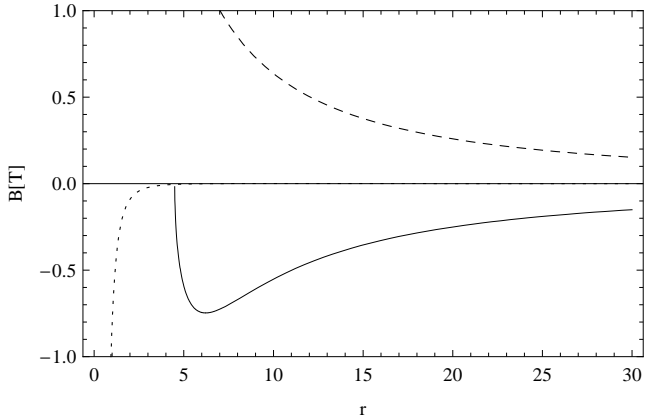}}
\caption{The toroidal magnetic field for the fiducial neutron star with a
spin period of 100\,s.  The solid and dashed lines show the field generated by
the dynamo for $\epsilon = -1$ and 1, respectively, while the dotted line shows
the magnetic field generated by the shear}
\label{toroidal2}
\end{figure}

We plot the magnetic fields for our fiducial model with varying $\epsilon$ and
spin periods of 7 and 100\,s, respectively,
in Figs. \ref{toroidal1} and \ref{toroidal2},
respectively. The corotation radius, at which $B_{\phi{\rm shear}}$  changes sign, occurs inside of
the inner edge of the accretion disc  for $\epsilon = 1$ and $-1$, and for these values of $\epsilon$,
$B_{\phi{\rm dyn}}$ is the dominant  magnetic field everywhere inside the disc.

In general the ratio of the magnetic field due to shear to that generated by the internal dynamo is
\begin{eqnarray}\label{57}
\frac{B_{\phi{\rm shear}}}{ B_{\phi{\rm dyn}}} = 6.25\times10^{-2}\epsilon^{-1}\gamma \gamma_{\rm dyn}^{-1/2}
\alpha_{\rm ss}^{-1/20} \bar{\mu}^{-3/16}M_1^{-11/56} \dot M_{13}^{2/35}\nonumber\\
\mu_{20}^{1/28}
\left(1-\omega_{s}r^{3/2}\right)\Lambda(r)^{-17/40}r^{-27/16}.
\end{eqnarray}
It is only during rather extreme conditions that the shear-induced field can
dominate at small radii, and the dynamo is always dominant at large radii.

\section{Discussion}

\subsection{The inner edge of the accretion disc}

We have shown that there are two different forms of inner disc boundaries that
can be found among our solutions.  We denote these as case D and V, respectively.
Case D occurs only for sufficiently negative $\epsilon$, while case V covers a
larger part of the parameter domain that we have studied.
We summarise our values for the inner radius of the accretion disc in
Tab.  \ref{tortab}.  We see that the inner radius can be significantly
larger than the Alfv\'en radius when $|\epsilon|$ is close to unity.
This is the result of the dynamo enhancing the coupling of the
stellar magnetic field to the accretion flow, such that it dominates
the viscous torque at larger radii than would otherwise be the case.

\subsection{The angular momentum balance}

To understand the exchange of angular momentum between the accretion
disc and its environment we multiply Eq. (\ref{19}) by $2\pi R$ and integrate
it from $R_0$, the inner radius of the disc, to $R_1$ the outer edge of the
disc
\begin{eqnarray}
  -\dot M \left(\sqrt{GMR_1}-\sqrt{GMR_0}\right)= \nonumber\\ \int_{R_0}^{R_1} 4\pi
\frac{B_z B_{\phi,{\rm dyn}}}{\mu_0} R^2 \mbox{d}R +
\int_{R_0}^{R_1} 4\pi \frac{B_z B_{\phi,{\rm shear}}}{\mu_0} R^2 \mbox{d}R\nonumber\\
-3\pi \left(\nu \Sigma\right)_{R_1} \sqrt{GMR_1}
+ 3\pi \left(\nu \Sigma\right)_{R_0} \sqrt{GMR_0}.
\end{eqnarray}
The lefthand side is the difference between the angular momentum that is
advected out of the inner edge of the accretion disc and that which is fed into
the disc at its outer edge and the righthand side describes the contribution of  magnetic and viscous torques
to the angular momentum balance.  The term $3\pi(\nu \Sigma)_{R_1}
\sqrt{GMR_1}$ describes the viscous tension at the outer edge of the
disc, which is not considered further in this paper.  Rather, we 
consider the exchange of angular momentum between the accretion disc and
the neutron star.  Firstly, we have the angular momentum which is
advected from the disc to the star
\begin{equation}\label{58}
    N_{adv} = \dot{M}\sqrt{GMR_{0}}= 2.6\times10^{26}\mu_{20}^{2/7}M_{1}^{3/7}
\dot{M}_{13}^{6/7}r_{0}^{1/2}.
\end{equation}
Secondly, we have the magnetic torques on the neutron star, which we divide into
one part due to the shear
\begin{eqnarray}\label{59}
    N_{\rm shr} = - \int_{R_0}^{R_1} 4\pi
  \frac{B_z B_{\phi,{\rm shear}}}{\mu_0} R^2 \mbox{d}R = \nonumber\\
  - 7.5\times 10^{26}
  \gamma \mu_{20}^{2/7}M_{1}^{3/7}\dot{M}_{13}^{6/7}\nonumber\\
  \int_{r_{0}}^{\infty}\left[r^{-4}\left(1 - \omega_{s}r^{3/2}\right)\right]dr
\end{eqnarray}
and a second part due to the dynamo
\begin{eqnarray}\label{60}
  N_{\rm dyn}= - \int_{R_0}^{R_1} 4\pi
  \frac{B_z B_{\phi,{\rm dyn}}}{\mu_0} R^2 \mbox{d}R =\nonumber\\ 
  1.2\times 10^{28}\epsilon\gamma_{dyn}^{1/2}\alpha_\mathrm{ss}^{1/20}
  \bar{\mu}^{3/16}\mu_{20}^{1/4}M_{1}^{5/8}\dot{M}_{13}^{4/5}\nonumber\\
  \int_{r_{0}}^{\infty}\Lambda^{17/40}r^{-37/16}dr.
\end{eqnarray}
Finally the viscous stress at the inner edge of the accretion disc transports
angular momentum outwards away from the neutron star resulting in a torque
\begin{eqnarray}\label{61}
    N_{\rm vis} = - 3\pi \left(\nu \Sigma\right)_{R_0} \sqrt{GMR_0} =\nonumber\\
-2.46\times10^{27}\mu_{20}^{2/7}M_{1}^{3/7}\dot{M}_{13}^{6/7}\Lambda r_{0}^{1/2}.
\end{eqnarray}
This torque vanishes for a case D inner boundary.
We can now calculate the torques on our fiducial neutron star for our choices of
spin periods and $\epsilon$.  These results are summarised in Table \ref{tortab}.
Unless  $\epsilon = 0$ we see that the magnetic torque due to the dynamo is always
significantly stronger than the magnetic torque due to the shear, and both are stronger for
$\epsilon = 0.1$ than for $\epsilon = 1$. The reason for this effect is that the central hole
in the disc grows too large  when $\epsilon = 1$.  The dominant torque at $\epsilon
= 1$ is therefore the viscous torque at the inner boundary, which has usually
been ignored.

Let us now compare our results with the BATSE data (\cite{bildsten}).
The 7.6\,s X-ray pulsar \object{4U1626-67} was observed to spin down at a rate
$\dot \nu \approx -7\times10^{-13}\,\rm Hzs^{-1}$ and spin up at
$\dot \nu \approx +8.5\times10^{-13}\,\rm Hz\,s^{-1}$.  These spin
variations correspond to torques
\begin{equation}
  N = 2\pi \dot \nu I = 6.3\times 10^{25} \dot \nu_{-13} I_{38}\,\mbox{Nm},
\end{equation}
where $I$ is the moment of inertia of the neutron star, which we measure in
$10^{38}$\,kg\,m$^2$, and we measure the spin acceleration in units of
$10^{-13}$\,Hz\,s$^{-1}$.  The required torque is significantly larger than what
is produced by
$N_{\rm shear}$ in any of our  models, but $N_{\rm dyn}$ is even an order of magnitude greater than
this value if $|\epsilon|= 0.1$. This overestimate can be explained by that we have assumed a unidirectional
$B_{\rm\phi,{\rm dyn}}$ across the disc surface, while it might be more realistic to expect that the toroidal
field is organised in magnetic annuli with opposite polarities, or that 
$|\epsilon|$ is significantly smaller fraction of the total turbulent magnetic 
field.

We explain the torque reversals in the same way as in Torkelsson (1998), where
the dynamo undergoes a field
reversal, which in our model corresponds to $\epsilon$ changing sign. One 
might speculate that, during this field reversal, the disc passes through a state 
corresponding to $\epsilon = 0$, in which the disc will have a smaller inner
radius than during the states with an active dynamo. The closer to the surface
of the neutron star that the accretion disc extends,
the more of the X-ray emission from the accretion columns it can absorb and
re-process.
Torkelsson (1998) showed that the observed time scales on which the torques
remain constant in the sources \object{Cen X-3}, \object{OAO 1657-415}, and 
\object{GX 1+4} are comparable
to the global viscous time scales of their accretion discs, which constrains
the mechanism that is responsible for the reversals of the magnetic field.

\begin{table*}
\caption{The inner edge of the accretion disc and its torque on the
fiducial neutron star.}
\label{tortab}
\centering
\begin{tabular}{lllllllll}
\hline
$P_{\rm spin}$\,[s] & $\epsilon$ & Case & $R_{0}$ & $N_{\rm shear}$ & $N_{\rm dyn}$ & $N_{\rm adv}$ & $N_{\rm vis}$ & $N_{\rm tot}$ \\
\hline
  7 & 1 &V& $9.0R_{A}$ & $7.0\times10^{24}$ &  $5.7\times10^{26}$ & $7.8\times10^{26}$ & $-1.2\times10^{27}$ & $2.6 \times10^{26}$\\
   & 0.1 & V&$2.0R_{A}$ &  $5.3\times10^{25}$ & $4.3\times10^{27}$ & $3.7\times10^{26}$ & $-4.9\times10^{26}$ & $4.2\times10^{27}$ \\
   & 0 & V&$1.0 R_{A}$ &  $5.9\times10^{25}$ & 0 & $2.6\times10^{26}$ &$-1.2 \times10^{26}$ & $2.0\times10^{26}$\\
   & -0.1 &D& $1.6R_{A}$ &  $6.3\times10^{25}$ & $-3.7\times10^{27}$ & $3.3\times10^{26}$ & 0 & $-3.3\times10^{27}$ \\
   & -1 & D&$4.7R_{A}$ & $1.8\times10^{25}$ & $-9.1\times10^{26}$ & $5.6\times10^{26}$ & 0 & $-3.2\times10^{26}$ \\
  18.7 & 1 & V&$9R_{A}$ &  $2.4\times10^{24}$ & $5.7\times10^{26}$ & $7.8\times10^{26}$ & $-1.1\times10^{27}$ & $2.5\times10^{26}$\\
   & 0.1& V&$2R_{A}$ &  $8.2\times10^{24}$ & $4.4\times10^{27}$ & $3.7\times10^{26}$ & $-5.4\times10^{26}$ & $4.2\times10^{27}$ \\
   & 0 & V&$1.0R_{A}$ &  $-6.7\times10^{25}$ & 0 & $2.6\times10^{26}$ & $-3.0\times10^{26}$ &$1.7\times10^{26}$ \\
   & -0.1 &V& $1.0R_{A}$ &  $-6.7\times10^{25}$ & $6.0\times10^{27}$ & $2.6\times10^{26}$ & $-2.5\times10^{25}$ & $-5.9\times10^{27}$ \\
   & -1 &D& $4.6R_{A}$ &  $6.0\times10^{24}$ & $-9.3\times10^{26}$ & $5.6\times10^{26}$ & $0$ &$-3.7\times10^{26}$ \\
  100 & 1 &V& $10R_{A}$ &  $2.9\times10^{23}$ &$4.9\times10^{26}$ & $8.2\times10^{26}$ & $-1.0\times10^{27}$ & $2.8\times10^{26}$ \\
   & 0.1 &V& $2.5R_{A}$ &  $-5.4\times10^{24}$ & $3.2\times10^{27}$& $4.1\times10^{26}$ &$-6.2\times10^{26}$ & $3.0\times10^{27}$ \\
   & 0 &V&$1.2R_{A}$ &  $-7.1\times10^{25}$ & 0 & $2.9\times10^{26}$ & $-4.4\times10^{26}$ & $-2.3\times10^{26}$ \\
   & -0.1 &V& $1.0R_{A}$ &  $-1.3\times10^{26}$ & $-7.1\times10^{27}$ & $2.6\times10^{26}$ & $-4.5\times10^{25}$ & $-6.9\times10^{27}$ \\
   & -1 &D& $4.4R_{A}$ & $-1.4\times10^{23}$ & $-9.4\times10^{26}$ & $5.5\times10^{26}$ & $0$ &$-4.0\times10^{26}$\\
\hline \\
\end{tabular}
\end{table*}

\section{Conclusions}

We have investigated the interaction between a magnetic neutron star and its
surrounding accretion disc in the case where the accretion disc is supporting
an internal dynamo.  The magnetic field that is produced by the dynamo can lead
to a significant enhancement of the magnetic torque between the neutron star
and the accretion disc, compared to what is seen in the model by
Ghosh \& Lamb (\cite{ghosh}).  
This extra magnetic torque
can explain the large variations in spin frequency \object{Cen X-3} and 
\object{OAO 1657-415} (Bildsten et al.1997).
Furthermore, a reversal of  the magnetic field that is generated by the dynamo, similar to the reversals
that we see of the magnetic fields on the Sun, could explain the torque reversals in these objects.

From the way that we calculate the structure of the accretion disc, we
find two kinds of solutions with different behaviours at the inner edge.
A few of our solutions have case D boundaries at which the density and
temperature go to 0 at finite radius, while most of our solutions have case V boundaries
at which the accretion is driven entirely by the magnetic tension between
the accreting matter and the neutron star.  In this case there is a
viscous stress between the accretion disc and the boundary layer, which
can transfer angular momentum outwards at a rate that is comparable to the
one at which it is advected inwards by the accreting matter itself.

We have also found that the dynamo leads to that the inner edge of the
accretion disc occurs at a radius that is larger than the traditional
Alfv\'en radius.  This effect is weak, though, for a realistic value of the dynamo-generated magnetic
field.

\begin{acknowledgements}
SBT thanks the Department of Physics at the University of Gothenburg for
hospitality and support during this project. SBT is supported in part by the
Swedish Institute (SI) under their Guest Scholarship Programme.  UT thanks
the Department of Physics at Addis Ababa University for their hospitality.
This research has made use of NASA's Astrophysics Data System.
We thank an anonymous referee for comments that have improved the
quality of the paper.
\end{acknowledgements}

\end{document}